\documentclass[aps,prl,reprint,nofootinbib,superscriptaddress,showpacs]{revtex4-1}

\usepackage{amsmath,amssymb,amsfonts}
\usepackage[utf8]{inputenc}
\usepackage{hyperref}

\newcommand{\rpctp}{Rudolf Peierls Centre for Theoretical Physics, University of Oxford}
\newcommand{\cecs}{Centro de Estudios Científicos (CECs) Av.~Arturo Prat~514, Valdivia, Chile}
\newcommand{\uab}{Universidad Andrés Bello, Av.~República~440, Santiago, Chile}
\newcommand{\ucsc}{Departamento de Matemática y Física Aplicadas, Universidad Católica de la Santísima Concepción, Concepción, Chile}
\newcommand{\udec}{Departamento de Física, Universidad de Concepción, Casilla 160-C, Concepción, Chile}
\newcommand{\ulbisi}{Physique Théorique et Mathématique, Université Libre de Bruxelles and International Solvay Institutes, Campus Plaine C.P.~231, B-1050 Bruxelles, Belgium}

\begin{document}

\title{The BTZ black hole as a Lorentz-flat geometry}

\author{Pedro D. Alvarez}
\email{alvarez@physics.ox.ac.uk}
\affiliation{\rpctp}

\author{Pablo Pais}
\email{pais@cecs.cl}
\affiliation{\cecs}
\affiliation{\uab}

\author{Eduardo Rodríguez}
\email{edurodriguez@ucsc.cl}
\affiliation{\ucsc}

\author{Patricio Salgado-Rebolledo}
\email{pasalgado@udec.cl}
\affiliation{\cecs}
\affiliation{\udec}
\affiliation{\ulbisi}

\author{Jorge Zanelli}
\email{z@cecs.cl}
\affiliation{\cecs}
\affiliation{\uab}

\begin{abstract}
It is shown that 2+1 dimensional anti-de Sitter spacetimes are Lorentz-flat. This means, in particular, that any simply-connected patch of the BTZ black hole solution can be endowed with a Lorentz connection that is locally pure gauge. The result can be naturally extended to a wider class of black hole geometries and point particles in three-dimensional spacetime.

\end{abstract}
\maketitle

Since its discovery, the Ba\~nados-Teitelboim-Zanelli (BTZ) black hole solution in three-dimensional spacetime \cite{BTZ} has been a source of surprise. The BTZ black hole shares all the features of the more realistic $3+1$ counterparts, such as the existence of an event horizon that surrounds the central singularity, its formation by collapsing matter, the emission of Hawking radiation consistent with thermodynamics, and the relation between entropy and the area of the horizon, among others. On the other hand, the enormous simplification resulting from the absence of propagating degrees of freedom in $2+1$ dimensions makes it an ideal laboratory to test gravitation theory in a lighter setting \cite{Carlip}.

In this note we show that the geometry of this spacetime has another exceptional feature: any simply connected patch $U$ of it is parallelizable with respect to a Lorentz connection. This means that $U$ can be covered with a family of locally inertial frames (freely falling observers) so that they can all be obtained by parallel transport from a given one $U_0$, independently of the path taken to connect them. The notion of parallelism here is the one relevant to the Lorentz group, characterized by the connection one-form $\omega^a{}_b= \omega^a{}_{b \mu} dx^\mu$. This connection defines the covariant derivative of a Lorentz vector $v^a$ with respect to the Lorentz group as \cite{footnote1}
\begin{equation}
Dv^a = dv^a + \omega^a{}_b v^b,
\end{equation}
and the corresponding Lorentz curvature is $R^a{}_b := d\omega^a{}_b + \omega^a{}_c \omega^c{}_b$.

The  geometry of the BTZ solution is the quotient of a constant negative curvature manifold (AdS$_3$) by an isometry that identifies points along a Killing vector \cite{BHTZ},
\begin{equation}
\mathcal{M}_{\text{BTZ}}=AdS_3 / \Gamma_K\, . \label{geometry}
\end{equation}
The metric of the spacetime is $ ds^2=-f^2dt^2 + f^{-2} dr^2 + r^2(N_\phi dt + d\phi)^2$, where $f^2 = -M+ r^2/\ell^2 + J^2/(2r)^2$, and $N_\phi = -J/(2r^2)$. Here $M$ is the mass, $J$ is the angular momentum, and the coordinates have the standard ranges, $-\infty <t<\infty$, $0<r<\infty$, $0\leq \phi \leq 2\pi$.  This is a solution of the Einstein equations obtained from the (2+1)-dimensional Einstein-Hilbert action with negative cosmological constant. Varying the action with respect to the metric, the field equations describe a manifold of constant negative Riemann curvature,
\begin{equation}
\mathcal{R}^{\alpha \beta}{}_{\mu \nu} = -\ell^{-2}\left(\delta^\alpha_\mu \delta^\beta_\nu - \delta^\alpha_\nu \delta^\beta_\mu \right)\, .
\end{equation}

As is well known, parallel transport of a vector (or a frame) in a closed loop produces a rotated vector (or frame) by a magnitude that depends on the total curvature enclosed by the loop. Hence, the possibility of covering the region $U$ with a family of parallel-transported frames independently of the path in a consistent manner, requires the corresponding curvature to vanish,
\begin{equation}
R^a{}_b (x) = 0\,, \quad \forall x \in U\, . \label{R=0}
\end{equation}

Since the Lorentz curvature does not make any reference to the metric $g_{\mu \nu}(x)$, a natural question to ask would be, what is the most general metric consistent with a Lorentz-flat geometry? In other words, does $R^{ab}=0$ determine, or impose some constraints, on the metric of the manifold? In order to answer this question, we can start by defining the metric in terms of the local frame one-forms (vielbeins), $e^a(x)=e^a{}_\mu(x) dx^\mu$,
\begin{equation}
g_{\mu \nu}(x) = \eta_{ab} e^a{}_\mu e^b{}_\nu\, . \label{g=ee}
\end{equation}
The vielbeins are vectors under the Lorentz group acting in the tangent space, and their covariant derivative defines the torsion two-form,
\begin{equation}
T^a=D e^a = de^a + \omega^a{}_b e^b\, ,  \label{T}
\end{equation}
which is also independent of the metric. However, the covariant derivative of the torsion must vanish, because $DT^a=R^a{}_b e^b$. So, we conclude that a Lorentz-flat spacetime $R^a{}_b=0$ can only admit a covariantly constant torsion,
\begin{equation}
DT^a = 0 \,.
\end{equation}
Splitting the Lorentz connection into a torsion-free part $\bar{\omega}^a{}_b$ and the \emph{contorsion}, $\kappa^a{}_b=\omega^a{}_b- \bar{\omega}^a{}_b$, we obtain
\begin{equation}
T^a =  \kappa^a{}_b e^b\,. \label{T-k}
\end{equation}
The Lorentz curvature can also be split into a purely metric part and torsion-dependent terms,
\begin{equation}\label{R-R-K}
R^a{}_b= \mathcal{R}^a{}_b + \bar{D} \kappa^a{}_b +\kappa^a{}_c\kappa^c{}_b\,,
\end{equation}
where $\mathcal{R}^a{}_b$ is the curvature for the torsion-free part of the connection, given by the Riemann tensor as
\begin{equation}
\mathcal{R}^{ab}=\frac{1}{2}e^a{}_\alpha e^b{}_\beta \mathcal{R}^{\alpha \beta}{}_{\mu \nu}dx^\mu dx^\nu. \label{Riem}
\end{equation}
In $2+1$ dimensions, the condition $DT^a =0$ can be integrated to
\begin{equation}\label{T2}
T^a = \tau \epsilon^a{}_{bc} e^b e^c \,, 
\end{equation}
where $\epsilon^a{}_{bc}=\eta^{ad}\epsilon_{dbc}$ is the Levi-Civita anti-symmetric invariant symbol. In (\ref{T2}) $\tau$ is a free integration parameter and therefore the value of the cosmological constant is not fixed and can take any non-positive value. From this last expression, the contorsion can be identified as $\kappa^a{}_b = - \tau \epsilon^a{}_{bc}e^c $. Using this expression in (\ref{R-R-K}) yields
\begin{equation}
R^{ab} =\mathcal{R}^{ab} + \tau^2 e^a e^b .  \label{R=ee}
\end{equation}
In other words, a spacetime with vanishing Lorentz curvature corresponds to an anti-de~Sitter ($\tau\neq 0$) or flat ($\tau= 0$) Riemannian geometry, where $\ell = 1/\tau$ is the radius of curvature. The origin of the sign in the cosmological constant can be easily traced to the Lorentzian signature. In fact, in a Euclidean space, the result (\ref{R=ee}) would have produced $\mathcal{R}^{ab}= \tau^2 e^a e^b$, which could be recognized as a result of the Adams-Hopf theorem \cite{Adams}: the three-sphere is parallelizable, namely, it can be endowed with a globally trivial $SO(3)$ connection. Equivalently, the statement that AdS$_3$ is Lorentz-flat is just the continuation to Lorentzian signature of the Adams-Hopf result. Since the Adams-Hopf theorem establishes that $S^7$ is parallelizable, one should expect that some interesting  covariantly constant torsion geometries would also exist in AdS$_7$.

Now, since 2+1 black holes for any $M$ and $J$ are obtained by an identification of AdS$_3$, all of them are locally Lorentz-flat. In fact, this feature can also be extended to other locally AdS$_3$ solutions, like the naked singularities obtained by identifications that produce a conical singularity \cite{DJT} and that correspond to the negative mass spectrum of the BTZ solution, $-1<M<0$.

Other local Lorentz flat black hole solutions can be constructed in the presence a locally flat but globally nontrivial gauge connection. This is the case, for instance in the vacuum sector of some supersymmetric Chern-Simons theories that include the $U(1)$ \cite{AVZ,APZ}, or $SU(2)$ \cite{APRSZ} connections. Those solutions, for particular values of the parameters, are configurations admitting gobally defined Killing spinors and therefore define stable BPS ground states.

Discussions with J. Gegenberg, G.Giribet, J. Gomis, A. Maloney and C. Martínez are gratefully acknowledged. P.P. is supported by grants from Universidad Andr{\'e}s Bello's Vicerrectorr{\'i}a de Investigaci{\'o}n y Doctorado.  P. S-R. is supported by grants from Becas Chile, CONICYT. P. D. A. is supported by by Beca Chile 74130061. Partial support by Universidad de Concepci\'{o}n, Chile is also acknowledged. This work was also partially supported by Fondecyt grant 1140155. The Centro de Estudios Cient\'{\i}ficos (CECS) is funded by the Chilean Government through the Centers of Excellence Base Financing Program of Conicyt.


\begin{thebibliography}{99}
\bibitem{footnote1} Here exterior (wedge) product of forms is implicitly assumed throughout.
\end{thebibliography}
\end{document}